\documentclass[twocolumn,epsfig,pre,showpacs]{revtex4}
\usepackage{graphicx}
\usepackage{dcolumn}
\usepackage{amsmath}    
\usepackage{amssymb}
\usepackage{bm} 
\usepackage{hyperref}
\usepackage{latexsym}
\usepackage{verbatim}
\usepackage{color}
\usepackage{subfigure}
\def\beq{\begin{equation}}
\def\eeq{\end{equation}}
\def\n{\hat{\bf n}}
\def\rb{{\bf r}}
\def\qb{{\bf q}}
\def\etal{{\it et al.}}
\def\sq{S_{\bf q}}
\def\intr{\int \mbox{d}^dr}

\def\K{\mbox{\sffamily\bfseries K}}

\def\nabbold{\mbox{\boldmath $\nabla$\unboldmath}}

\def\beq{\begin{equation}}                           
\def\eeq{\end{equation}}                           
\def\bea{\begin{eqnarray}}                           
\def\eea{\end{eqnarray}}        

\begin{document}
\textwidth = 6.5 in
\textheight = 9 in
\oddsidemargin = 0.0 in
\evensidemargin = 0.0 in
\topmargin = 0.0 in
\headheight = 0.0 in
\headsep = 0.0 in
\parskip = 0.2in
\parindent = 0.0in

\title{Active nematics are intrinsically phase-separated}
\author{Shradha Mishra}
\author{Sriram Ramaswamy\footnote{Also with CMTU, JNCASR, Bangalore 560064, India}}
\email{sriram@physics.iisc.ernet.in}
\affiliation{Centre for Condensed Matter Theory, Department of Physics,
Indian Institute of Science, Bangalore 560 012 INDIA}
\date{accepted for publication in Phys Rev Lett 3 Aug 06}
\begin{abstract}
Two-dimensional nonequilibrium nematic steady states, as found in 
agitated granular-rod monolayers or films of orientable amoeboid
cells, 
were predicted [Europhys. Lett. {\bf 62} (2003) 196] 
to have giant number fluctuations, with standard
deviation proportional to the mean. 
We show numerically that the steady state of such systems is  
{\em macroscopically phase-separated}, yet dominated by fluctuations, 
as in the Das-Barma model [PRL {\bf 85} (2000) 1602].  
We suggest experimental tests of our findings in granular and living-cell systems. 
\end{abstract} 
\maketitle

The ordering or ``flocking'' \cite{vicsek,tonertu,tonertusr} 
of self-propelled particles obeys laws strikingly
different from those governing thermal equilibrium systems of the same spatial
symmetry. Even in two dimensions, the velocities of particles 
in such flocks show true long-range order \cite{vicsek,tonertu}, 
despite the spontaneous breaking of continuous rotational invariance.
Density fluctuations in the ordered phase are 
anomalously large \cite{tonertu}, and the onset of the ordered phase is 
discontinuous \cite{gregoirechate}. The ultimate origin of these nonequilibrium 
phenomena is that the order parameter is not simply an orientation but a 
macroscopic velocity. 
It is thus intriguing that even the {\em nematic} phase of 
a collection of self-driven particles, which is apolar and 
hence has zero macroscopic velocity, shows 
\cite{sradititoner,chateginellimontagne} giant number fluctuations \cite{other},  
as a result of the manner in which orientational fluctuations drive mass currents.
This Letter takes a closer look at these fluctuations 
and shows that they offer a physical realisation of
the remarkable nonequilibrium phenomenon known as {\em fluctuation-dominated 
phase separation} \cite{dasbarma}, hitherto a
theoretical curiosity. 


Before presenting our results, we make 
precise the term {\em active nematic}. 
An {\em active particle} extracts energy from sources 
in the ambient medium or an internal fuel tank, 
dissipates it by the  
cyclical motion of an internal ``motor'' coordinate, and moves as a consequence. 
For the anisotropic particles that concern us here, the 
direction of motion is determined predominantly by the orientation. 
Our definition encompasses 
self-propelled organisms, living cells, molecular motors, 
and macroscopic rods on a vertically vibrated substrate (where 
the tilt of the rod serves as the motor coordinate).  
An {\em active nematic} is a collection of such particles with 
axes on average spontaneously aligned in a direction $\hat{\bf n}$, 
with invariance under $\hat{\bf n} \to -\hat{\bf n}$.   
We know of two realisations of active nematics: collections of 
living amoeboid cells \cite{gruler} and   
granular-rod monolayers \cite{kudrolli,vjmenonsr}.  

%
%
\begin{figure}[htbp]
\begin{center}
{\includegraphics[width=6.0cm]{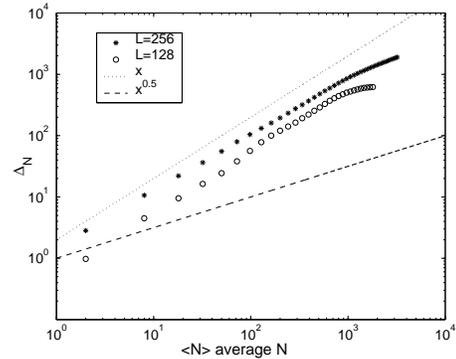}}
\caption{Number standard deviation $\Delta_N$ scales roughly as the mean $\bar{N}$, 
for system sizes $L=128, 256$}
\label{standdev}
\end{center}
\end{figure}

\begin{figure}[htbp]
\begin{center}
{\includegraphics[width=6.0cm]{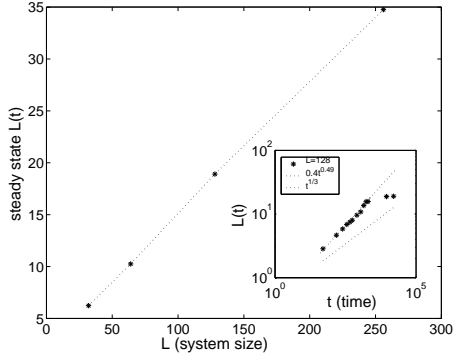}}
\caption{Coarsening length $L(t)$ saturates to a value proportional 
to system size. Inset: $L(t)$ consistent with $t^{1/2}$ despite 
conservation law. } 
\label{charlengt}
\end{center} 
\end{figure}
\begin{figure}[htbp]
\begin{center}
\includegraphics[width=6.0cm]{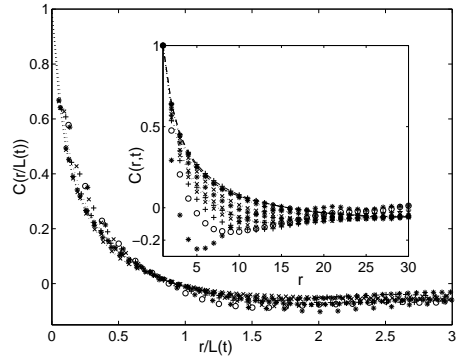}
\caption{Plots of the equal-time density
correlator $C(r,t)$ at different times $t$ collapse onto a 
single curve when $r$ is scaled by the zero-crossing coarsening length $L(t)$;  
data shown for direction transverse to nematic ordering
Inset: $C(r,t)$ {\it vs.} $r$.}
\label{densitycorre}
\end{center}
\end{figure}
\begin{figure}[htbp]
\begin{center}
\includegraphics[width=6.0cm]{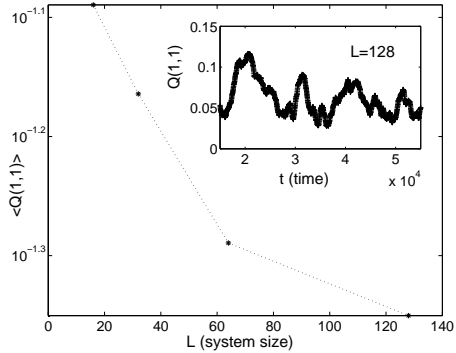}
\caption{Time-averaged fourier component of the density profile 
 {\it vs.} wave vector $k$ for wave vector
direction (1,1), for system sizes {\em $L=16, 32, 64, 128$}; flattening in 
semilog plot at largest size suggests nonzero value for $L \to \infty$. 
Inset: Time series of $Q$ for (1,1) direction.
}
\label{fourierdensity}
\end{center}
\end{figure}

We study active nematics in a simple numerical model described in detail 
below. Our results confirm (see Fig. \ref{standdev}) the giant number 
fluctuations (standard deviation $\propto$ mean)  
\cite{chateginellimontagne} 
predicted by the linearised analysis of \cite{sradititoner}, but are far richer:  
(i) A statistically uniform initial distribution of particles, on 
a well-ordered nematic background, undergoes a delicate 
``fluctuation-dominated'' \cite{dasbarma} phase separation, 
where the system explores many statistically similar segregated 
configurations. 
(ii) The equal-time two-point density correlator $C(\rb)$, Fig. \ref{densitycorre},  
shows a collapse 
when plotted as a function of $r/L(t)$, where $L(t)$ is the 
location of the first zero-crossing of $C(\rb)$, with a cusp at small $r/L(t)$ 
signalling a departure from Porod's Law, i.e., the absence of sharp  
interfaces. 
(iii)  We confirm that the large density inhomogeneities are 
indeed best thought of as phase separation, by showing (1) that 
the saturation value of $L(t)$ is proportional to the linear size 
of the system and (2) that the 
phase-separation order parameter -- the time-averaged first spatial 
Fourier component of the particle density -- approaches a nonzero value 
in the limit of large system size (Fig. \ref{fourierdensity}).  
(iv) $L(t)$ grows clearly faster than the $t^{1/3}$ for normal 
conserved-order-parameter coarsening (see inset to Fig. \ref{charlengt}),  
and is consistent with $t^{1/2}$ as expected by analogy to \cite{dasbarma}. 
Below we show how these results were obtained, discuss them  
in detail, explain the analogy to the work of \cite{dasbarma}, and suggest experimental 
tests for these striking phenomena.

We begin with some background information. An apolar, 
uniaxial, compressible nematic liquid crystal is described by 
director and number-density  
fields $\n(\rb)$ and $c(\rb)$, with 
fluctuations $\delta {\bf n}(\rb)$ and $\delta c(\rb)$ about their uniform 
mean values $\n_0$ and $c_0$. Let us review first 
what happens at {\em thermal equilibrium}. The system 
is then governed by an extended Frank \cite{degp} 
free-energy $F[\n,c] =  (1/2)\intr[\K(\nabbold \n)^2 
+A(\delta c)^2/c_0 + C_1 \n \cdot \nabla c \nabla \cdot \n  
+ C_2 \n \times \nabla c \cdot \nabla \times \n]$, 
where $\K$ is an elastic tensor, $A$ the compressional 
modulus at constant orientation, and $C_{1,2}$ couple 
orientation and density in the simplest symmetry-allowed fashion.  
Equipartition applied to $F$ implies 
that the static structure factor $\sq 
\equiv  
\intr \exp(-i \qb \cdot \rb)\langle \delta c ({\bf 0}) \delta c (\rb) \rangle /c_0$  
is {\em finite} for $q \to 0$; i.e., the mean $\bar{N}$ and 
standard deviation $\Delta_N$ of the number $N$ of particles  
obey $\Delta_N \propto \sqrt{\bar{N}}$ at equilibrium even when $C_1,\,C_2 \neq 0$. 

\begin{figure} 
\begin{center}
\includegraphics[width=1.0in]{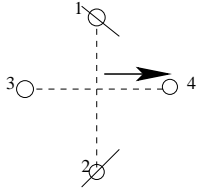}
\caption{Director variation from the bottom to the top of the picture gives
rise to a current in a transverse direction } 
\label{curvcoupling} 
\end{center}
\end{figure}

An active nematic is a steady state away from thermal equilibrium,  
in which not only dynamic correlators but equal-time quantities like 
$\sq$ or $\Delta_N$ as well must be inferred from equations of motion 
for $\n$ and $c$. 
As shown in \cite{sradititoner}, the equation of motion 
for $\n$ is qualitatively the same as for equilibrium nematics. 
The feature \cite{sradititoner} that distinguishes active nematics crucially from 
their equilibrium counterparts is that the current 
${\bf j}$ in the continuity equation 
$\partial_t c = -\nabla \cdot {\bf j}$ for the density  
has a contribution $\propto \nabla \cdot c (\n \n)$ 
\cite{ahmadietallong}. 
This term, which is ruled out at thermal 
equilibrium, has a simple, physically appealing origin:
spatial variation in the director field $\n$ defines a curve; 
the normal to this curve defines a local vectorial asymmetry;  
for a driven system, such an asymmetry implies a current \cite{curie}. 
If $\n = (\cos \theta, \sin \theta)$ then inhomogeneities in $\theta$ 
give a curvature-induced current 
${\bf j} = (j_x, j_z) 
\propto (\partial \theta / \partial z, \partial \theta / \partial x)$ 
(see Fig. \ref{curvcoupling}), analogous to \cite{dasbarma} where 
particles slide with velocity $\propto \nabla h$ 
on a fluctuating interface with height field $h$, except that 
our current is not a gradient. the  Since nematic order is a spontaneous breaking 
of rotation-invariance, large fluctuations in $\theta$ 
at long wavelengths are  
expected to be present in abundance, and to decay slowly, 
in any nematic, equilibrium or otherwise. We showed in the previous 
paragraph that the effect of these broken-symmetry modes on 
the density field was benign in an {\em equilibrium} nematic.
In an {\em active} nematic, however, the same orientational fluctuations, 
because of the curvature-induced current 
we just mentioned, will affect the density fluctuations substantially.
A linearised small-fluctuations 
analysis \cite{sradititoner} showed that they to lead to giant fluctuations in the number 
of particles: 
$\Delta_N/\sqrt{\bar{N}} \propto \bar{N}^{1/d}$ in $d$ dimensions, i.e., 
$\Delta_N \propto \bar{N}$ for $d = 2$. 

Such large fluctuations prompt the suspicion that an analysis beyond Gaussian
fluctuations would reveal that the system is in fact phase-separated, as in
\cite{dasbarma}. There are
two issues here: (i) whether the nonequilibrium coupling mentioned above
inevitably arises in an active nematic;  and (ii) whether it leads to 
phase separation. 
Ref. \cite{chateginellimontagne} effectively answers the first question in the
affirmative; 
we focus on the second. 

We find it convenient to separate the density and orientation degrees of
freedom, and employ a discrete model of lattice-gas particles coupled to an
angle field, incorporating explicitly the the nonequilibrium curvature-induced
current ${\bf j} \propto (\partial_z\theta,\partial_x\theta)$ mentioned above, 
via a suitable choice of particle-hopping rates.
We  consider a two-dimensional lattice with angles
$\theta_{i}$ $\epsilon$ $[0,\pi]$ and noninteracting lattice-gas occupancy 
variables $n_{i} = 0, 1$ at each site $i$. 
The angles evolve by Metropolis Monte Carlo updates
governed by the Lebwohl-Lasher \cite{leblash} hamiltonian $H=-K\sum_{<ij>} \cos
2(\theta_{i}-\theta_{j})$ yielding a nematic phase at low temperature.
Particle motion is nonequilibrium: Hops of a particle at site $i$ to a
nearest-neighbour site in the $\pm x$ direction are attempted with probability
$1/4  \pm \alpha (\theta_{1}-\theta_{2})$, and in the $\pm z$ direction with
probability  $1/4 \pm \alpha (\theta_{3}-\theta_{4})$, where $ \theta_{i}$ are
the angles at sites $i = 1 \mbox{ to } 4$ as in Fig. \ref{curvcoupling},
and $\alpha$ encodes the strength \cite{sign} of the active curvature-current coupling of
\cite{sradititoner}.  
\begin{figure}[htbp] 
\begin{center}
{\includegraphics[width=5.0cm]{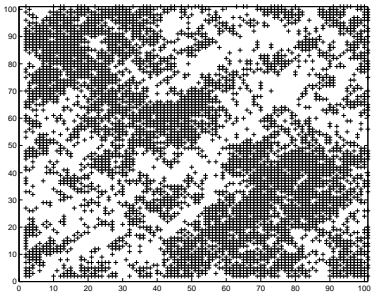}}\\ 
\caption{Anisotropic clustering of
particles, 20000 time steps,  system size 100, $50$ \% occupancy}
\label{realspace} 
\end{center} 
\end{figure} 
In \cite{sradititoner}, the effect
of the concentration field on the dynamics of the angle field is shown to play
an insignificant role in the giant number fluctuations. Accordingly, we neglect
it here, so that the particles are advected passively 
by an autonomous angle field. This simplifying
approximation should not make a qualitative difference deep in the nematic
phase.  Accordingly, we work with $K = 10.0$, and let the angle field
equilibrate until a well-ordered nematic is formed. 
An initial statistically uniform distribution of particles, (mean
occupancy =50 \%) clusters and coarsens (Fig. \ref{realspace}), most strongly at $\pm
45^{\circ}$ to the mean direction of nematic ordering.  The anisotropic
two-point density correlator $C(\bf{r},t)$ shows a scaling collapse for a given
direction for all $t$, if plotted as a function of $r/L(t)$ for a coarsening
length $L(t)$, defined by the value of $r$ at the first zero-crossing, whose
scaling is consistent with $t^{1/2}$.  (see Fig. \ref{densitycorre} and Fig.
\ref{charlengt}).  The value at which $L(t)$ saturates is proportional to the
system size $L$ (see Fig. \ref{charlengt}), which makes a strong case for true
phase separation. The value of the saturation length is numerically small
compared to $L$, probably because of the poorly-defined clusters (Fig.
\ref{realspace}) of fluctuation-dominated phase separation. That we work with
hard-core particles, and on timescales on which the macroscopic variation of
the mean nematic orientation is very small, also probably contributes.  The
exponent of $1/2$ is because the particles aggregate {\em not} by diffusion
plus short-range capture, but rather, by  analogy with \cite{dasbarma} 
(see also \cite{daspuri}), because
the broken-symmetry mode of the nematic order sweeps the particles over large
distances via curvature-induced drift.  A nematic fluctuation on a scale $\ell$
collects particles in a time of order $\ell$. For two such domains to 
coalesce requires the nematic director field on that scale to turn over, which
is a time $\sim \ell^z$ where $z = 2$ is the dynamic exponent of transverse
fluctuations of the nematic director.  

%

At long times a steady state is reached, 
and $C(r/L(t \to \infty))$ 
shows a cusp at
small argument ($C(x)\propto x^{a}, a \simeq 0.33)$, 
which can be seen in Fig. \ref{densitycorre} as well, signalling the
absence of sharp interfaces between regions rich and poor in particles,  
and a power-law distribution 
of cluster sizes. 
For steady state in the largest system, we also measured the standard deviation
$\Delta_{N}$ in the number of particles in an observation containing $\bar{N}$
particles on average. The plot of $\Delta_{N}$ vs $\bar{N}$, Fig. \ref{standdev}, 
shows precisely the
linear dependence predicted by \cite{sradititoner}.  
Faced with these results, we ask: is
this phase separation or a single phase with large fluctuations? This is
answered by measuring the magnitude of the time-averaged lowest spatial
Fourier-component $Q(1,1)$ of the density, shown in Fig. \ref{fourierdensity}. 
Although the data are not conclusive, the flattening of the semilog plot as a function 
of system size rules out an exponential decay to zero. Together with the 
proportionality of the coarsening length to the system size, and the 
nature of the mechanism, this suggests strongly that  
active nematics offer the most natural physical realisation of 
macroscopic fluctuation-dominated phase separation \cite{dasbarma}.  
As in \cite{dasbarma}, we find that the time-series of $Q(1,1)$ shows enormous
fluctuations, 
(Fig. \ref{fourierdensity}, inset), 
with ``crashes'' during which other nearby
fourier-components gain weight.  Thus, as in \cite{dasbarma}, 
the system lurches from one
macroscopically phase-separated configuration to another, spending very little
time in non-phase-separated states.  
Lastly, the velocity
autocorrelation of tagged particles at low (15 \%) concentration agrees  
qualitatively with the $1/t$ tail (plot not shown) 
predicted by  
\cite{sradititoner}, over the range in which a 
given particle moves unimpeded by others. 

What experiments can test these results?  The best would be
agitated layers of granular rods, for which nematic phases have been
reported \cite{vjmenonsr}. Although many features of 
these systems can be rationalized in terms of equilibrium hard-rod theories 
\cite{equihardrod}, some 
properties such as global circulation and swirls 
\cite{vjmenonsr,kudrolli} are clearly very nonequilibrium.  
These systems as well as the living melanocyte
nematic of Gruler {\etal} \cite{gruler} remain the most promising candidates 
for experimental tests 
of the rich range of results made here and in \cite{sradititoner}. 
The confirmation of giant number fluctuations in the numerical 
experiments of \cite{chateginellimontagne} is encouraging in this regard.  

The DST, India supports the Centre for Condensed Matter Theory.
SR thanks the IFCPAR (project 3504-2) and SM the CSIR, India for support. 
We acknowledge valuable discussions with M. Barma, S. Puri, 
C. Dasgupta and 
especially H. Chat\'{e}, who also generously shared his results 
\cite{chateginellimontagne} before publication. 

{}

\end{document}